\documentclass{shep3}

\usepackage{axodraw,epsfig}
\usepackage{cite}
  \usepackage{graphicx}
  \usepackage{amssymb}
  \usepackage{amsmath}

\preprint{
  LPSC 10/191 \\
  ZU-TH 18/10 \\
  BI-TP 2010/47
}

\title{Two-Loop Leading Color Corrections
to Heavy-Quark Pair Production in
the Gluon Fusion Channel}

\author{R.~Bonciani$\rm \, ^{a, \,}$\footnote{Email: {\tt
bonciani@lpsc.in2p3.fr}},  A.~Ferroglia$\rm \, ^{b,
\,}$\footnote{Email: {\tt
AFerroglia@citytech.cuny.edu}},  T.~Gehrmann$\rm \, ^{c,
\,}$\footnote{Email: {\tt thomas.gehrmann@uzh.ch}},
A.~von~Manteuffel$\rm \, ^{c,
\,}$\footnote{Email: {\tt manteuffel@physik.uzh.ch}}, and
C.~Studerus$\rm \, ^{d,\,}$\footnote{Email: {\tt cedric@physik.uni-bielefeld.de}} \\

\noindent {\it $\rm ^a$ Laboratoire de Physique Subatomique et de Cosmologie,
Universit\'e Joseph Fourier/CNRS-IN2P3/INPG,
F-38026 Grenoble, France}\\

\noindent {\it $\rm ^b$  New York City College of Technology
        300 Jay Street, NY 11201 Brooklyn, US }\\

\noindent {\it $\rm ^c$ Institut f{\"u}r
Theoretische Physik,
Universit{\"a}t Z\"urich,
CH-8057 Z\"urich, Switzerland}\\

\noindent {\it $\rm ^d$ Fakult{\"a}t f{\"u}r Physik,
Universit{\"a}t Bielefeld,
D-33501 Bielefeld, Germany }

}

\abstract{We evaluate the two-loop QCD diagrams contributing to the leading
color coefficient of the heavy-quark pair production cross section in the gluon 
fusion channel. We obtain an analytic expression, which is 
valid for any value of the Mandelstam 
invariants $s$ and $t$ and of the heavy-quark mass $m$. 
Our findings agree with previous analytic results in the small-mass limit and 
with recent results for the coefficients of the IR poles. }

\keywords{Heavy-Quark Production, Two-Loop Calculation}

\begin{document}

\newcommand{\be}{\begin{equation}}
\newcommand{\ee}{\end{equation}}
\newcommand{\bfm}[1]{\mbox{\boldmath$#1$}}
\newcommand{\bff}[1]{\mbox{\scriptsize\boldmath${#1}$}}
\newcommand{\al}{\alpha}
\newcommand{\bt}{\beta}
\newcommand{\lm}{\lambda}
\newcommand{\bea}{\begin{eqnarray}}
\newcommand{\eea}{\end{eqnarray}}
\newcommand{\gm}{\gamma}
\newcommand{\Gm}{\Gamma}
\newcommand{\dl}{\delta}
\newcommand{\Dl}{\Delta}
\newcommand{\ep}{\varepsilon}
\newcommand{\vep}{\varepsilon}
\newcommand{\kp}{\kappa}
\newcommand{\Lm}{\Lambda}
\newcommand{\om}{\omega}
\newcommand{\pa}{\partial}
\newcommand{\nn}{\nonumber}
\newcommand{\dd}{\mbox{d}}
\newcommand{\grtsim}{\mbox{\raisebox{-3pt}{$\stackrel{>}{\sim}$}}}
\newcommand{\lessim}{\mbox{\raisebox{-3pt}{$\stackrel{<}{\sim}$}}}
\newcommand{\uk}{\underline{k}}
\newcommand{\gsim}{\;\rlap{\lower 3.5 pt \hbox{$\mathchar \sim$}} \raise 1pt \hbox {$>$}\;}
\newcommand{\lsim}{\;\rlap{\lower 3.5 pt \hbox{$\mathchar \sim$}} \raise 1pt \hbox {$<$}\;}
\newcommand{\Li}{\mbox{Li}}
\newcommand{\bc}{\begin{center}}
\newcommand{\ec}{\end{center}}

\def\lapprox{\lower .7ex\hbox{$\;\stackrel{\textstyle <}{\sim}\;$}}
\def\gapprox{\lower .7ex\hbox{$\;\stackrel{\textstyle >}{\sim}\;$}}

%
\newcommand{\hypF}{{}_2\mbox{F}_1}


\section{Introduction}

The top quark is the heaviest elementary particle known to date. Its large mass  of
$173$ GeV makes it a crucial tool for the study of the electroweak symmetry breaking
mechanism and for the search of signals of physics beyond the Standard Model. The
study of the properties of the top quark at the Tevatron during the last 15 years
provided precise measurements~\cite{exptop} of the  top-quark mass (with a relative
error of $0.75 \%$) and of the total top-quark pair production cross section (with a
relative error of  about $10 \%$). Owing to the by-now large Tevatron dataset, 
measurements of differential distributions in top pair  production are becoming
possible~\cite{exp,Frederix:2010cn}.  A new set of precise experimental measurements
is expected to be obtained by the LHC experiments soon. At the LHC, running at 7~TeV,
with an integrated luminosity of only $\sim 200 \, \rm{pb}^{-1}$, it should be
possible to record about $30000$ top quark pair events before selection
\cite{Bernreuther:2010cv}. With the planned increases in center of mass energy and
luminosity, the LHC will turn into a veritable top-quark factory, producing millions
of top-quark events per year \cite{Bernreuther:2008ju}.  

In order to take full advantage of the improved experimental measurements, it will be
necessary to provide theoretical predictions for the measured observables which are
as accurate as possible. Most of the observables related to the top-quark pair
production have been calculated up to NLO  \cite{NLOcalc}. In several cases, the
next-to-leading (NLO) corrections have been supplemented by the resummation of large
logarithmic corrections at leading (LL, \cite{LL}),  next-to-leading (NLL,
\cite{NLL}) and next-to-next-to-leading logarithmic (NNLL,
\cite{NNLL,Ahrens:2010mj,Ferroglia:2010mi,Kidonakis:2010dk})  accuracy. However, to
match the  precision of the forthcoming experimental data, full
next-to-next-to-leading order (NNLO) calculations are required for at least some of
the observables, such as the top-quark  pair production total cross section
\cite{APPROX}.

To obtain NNLO predictions for the top quark pair production in QCD, the following
matrix elements need to be computed: {\em i)} two-loop matrix elements in both the
quark-antiquark annihilation channel and the gluon-fusion channel; {\em ii)} one-loop
matrix elements with  one extra parton in the final state; and {\em iii)} tree-level
matrix elements with two extra partons in the final state. The diagrams belonging to
the sets {\em ii)}  and {\em iii)} form part of the NLO corrections to
top-pair-plus-jet production and have been calculated by several groups 
\cite{Dittmaier:2007wz,Bevilacqua:2010ve,Melnikov:2010iu}. Aside from the calculation
of the required Feynman diagrams, a full NNLO calculation of the top-quark pair
production cross section is particularly challenging because it entails the
development of a NNLO subtraction method, which could be used in processes with
massive partons \cite{secdec,ant,cg,Czakon:2010td,Anastasiou:2010pw}. 

As far as matrix elements are concerned, the last missing ingredient to the NNLO
corrections to top quark pair production are the  two-loop matrix elements for $q\bar
q \to t\bar t$ and $gg\to t\bar t$. Both types of matrix elements were calculated in
the $s \gg m^2$ limit (where $s$ is the partonic center of mass energy and $m$ is the
mass of the top quark) \cite{Czakon:2007ej, Czakon:2007wk}. A fully numerical
calculation of the  two-loop corrections in the quark-antiquark annihilation channel
is also available \cite{Czakon:2008zk}. 

This paper is the third of a series of works aiming towards the analytic calculation
of the two-loop matrix elements for the top pair production process. For what
concerns the quark-antiquark annihilation channel, the two-loop diagrams involving a
closed light or heavy-quark loop were evaluated in \cite{quarkloops}, while the
two-loop diagrams contributing to the leading color coefficient were evaluated in
\cite{planarqq}. In both cases, the results obtained retain the full dependence on
the top-quark mass and on the kinematic invariants; they agree with the numerical
results of~\cite{Czakon:2008zk}. Having analytical results available has several
advantages over a purely numerical representation. Besides their considerably shorter
evaluation time, the analytical results also allow for an expansion in different
kinematical limits (threshold, high energy).

In the present paper, an analytical expression for the  two-loop diagrams
contributing to the leading color coefficient in the gluon-fusion channel is derived.
We carry out the calculation by employing the technique based on the Laporta
algorithm  \cite{Laportaalgorithm} and the differential equation method
\cite{DiffEq}, already used in \cite{quarkloops,planarqq}.  The calculation of the
leading color coefficient in the gluon fusion does not require the calculation of any
new master integrals beyond the ones obtained in the two previous works,  such that
we do not discuss the calculational method in full detail.  The interested reader can
find in \cite{quarkloops,planarqq} a detailed description of the  techniques
employed.

The paper is organized as follows: in Section~\ref{notconv}, we introduce our
notation and conventions; in Section~\ref{sec:calc}, we summarize the most relevant
features of our calculational method.  Section~\ref{sec:renorm} describes the UV
renormalization of the bare amplitude. The resulting two-loop amplitude
contributions are described in  Section~\ref{sec:results}, where  we also provide
numerical values in some benchmark points, and  discuss the expansion in the
threshold limit. Finally,  Section~\ref{sec:conc} contains our conclusions.


\begin{figure}
\vspace*{.5cm}
\[ 
\hspace*{1.5cm}
\vcenter{ \hbox{
  \begin{picture}(0,0)(0,0)
\SetScale{1}
  \SetWidth{.5}

\Gluon(-55,30)(-30,0){3}{8}
\Gluon(-55,-30)(-30,0){3}{8}
%
\Gluon(-30,0)(20,0){3}{10}
\LongArrow(-62,24)(-50,10)
\LongArrow(-62,-24)(-50,-10)
\LongArrow(40,10)(52,24)
\LongArrow(40,-10)(52,-24)
  \SetWidth{1.6}
\ArrowLine(45,-30)(20,0)
\ArrowLine(20,0)(45,30)

\Text(-67,27)[cb]{$p_1$}
\Text(-67,-33)[cb]{$p_2$}
\Text(59,-33)[cb]{$p_4$}
\Text(59,27)[cb]{$p_3$}
\Text(-2,-55)[cb]{(a)}
\end{picture}}
}
\hspace*{5.5cm}
\vcenter{ \hbox{
  \begin{picture}(0,0)(0,0)
\SetScale{1}
  \SetWidth{.5}

\Gluon(-55,30)(-15,30){3}{8}
\Gluon(-55,-30)(-15,-30){3}{8}
  \SetWidth{1.6}
\ArrowLine(-15,30)(25,30)
\ArrowLine(25,-30)(-15,-30)
\ArrowLine(-15,-30)(-15,30)
\Text(-15,-55)[cb]{(b)}

\end{picture}}
}
\hspace*{4cm}
\vcenter{ \hbox{
  \begin{picture}(0,0)(0,0)
\SetScale{1}
  \SetWidth{.5}

\Gluon(-55,30)(-15,30){3}{8}
\Gluon(-55,-30)(-15,-30){3}{8}
  \SetWidth{1.6}
\Line(-15,30)(3,3)
\ArrowLine(25,-30)(7,-4)
\Line(-15,-30)(5,0)
\ArrowLine(5,0)(25,30)
\ArrowLine(-15,30)(-15,-30)
\Text(-15,-55)[cb]{(c)}

\end{picture}}
}
\]
\vspace*{.8cm} 
\caption{\it Tree-level  amplitude. Massive
quarks are indicated by a thick line.} 
\label{figTree}
\end{figure}
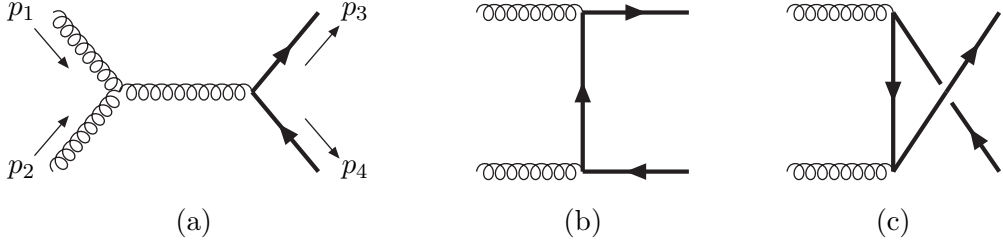


\section{Notation and Conventions \label{notconv}}

We consider the scattering process \be g(p_1) + g(p_2)
\longrightarrow  t(p_3) + \bar{t}(p_4) \, , \ee
in Euclidean kinematics, where $p_i^2 = 0$ for $i=1,2$  and
$p_j^2 = -m^2$ for  $j=3,4$. The Mandelstam variables are defined as
follows
\be
s = -\left(p_1 + p_2 \right)^2 \, , \quad
t = -\left(p_1 - p_3 \right)^2 \, , \quad
u = -\left(p_1 - p_4 \right)^2 \, .
\ee
Conservation of momentum implies that $s +t +u = 2 m^2$.

The squared matrix element (summed over spin and color),  calculated in $d = 4 -2
\varepsilon$ dimensions, can be expanded in powers  of the strong coupling constant
$\alpha_S$ as follows:
\be \label{M2}
\sum |\mathcal{M}|^2(s,t,m,\varepsilon) = 16 \pi^2 \alpha_S^2
\left[{\mathcal A}_0 +
\left(\frac{\alpha_s}{ \pi} \right) {\mathcal A}_1 +
\left(\frac{\alpha_s}{ \pi} \right)^2 {\mathcal A}_2 +
{\mathcal O}\left( \alpha_s^3\right)\right] \, .
\ee
The tree-level amplitude involves the three diagrams shown in 
Fig.~\ref{figTree} and their contribution to Eq.~(\ref{M2}) is given by
\bea
\hspace*{-5mm} 
{\mathcal A}_0 & = &  (N_c^2-1) \Biggl\{ \left[
       \frac{1}{N_c}   \frac{1}{t_1^2 u_1^2}
       - N_c  \, \frac{(t_1^2+u_1^2)}{t_1^2 u_1^2
   (t_1+u_1)^2} \right] F_1(t_1,u_1,m^2)
       + \varepsilon \biggl[
       \frac{1}{N_c}  \frac{4 \left(t_1^2+u_1
   t_1+u_1^2\right)}{t_1 u_1} \nn\\
& &
       - N_c \frac{4 \left(t_1^2+u_1^2\right)
   \left(t_1^2+u_1
   t_1+u_1^2\right)}{t_1 u_1
   (t_1+u_1)^2}
            \biggr]
       - \varepsilon^2 \biggl[
         \frac{1}{N_c}  \frac{2(t_1+u_1)^2}{t_1 u_1}
       - N_c \frac{2(t_1^2+u_1^2)}{t_1 u_1}
            \biggr] \Biggr\} ,
\label{treeM}
\eea
where  $N_c$ is the number of colors, $t_1=t-m^2$, $u_1=u-m^2$, and  
\be
F_1(t_1,u_1,m) =
 - 2 t_1 u_1 (t_1^2 + u_1^2)
 + 8 m^2 t_1 u_1 (t_1 + u_1)
 + 8 m^4 (t_1 + u_1)^2 \,.
\ee

The ${\mathcal O}(\alpha_S)$ term ${\mathcal A}_1$  in Eq.~(\ref{M2}) arises  from
the interference of one-loop diagrams with the tree-level amplitude \cite{NLOcalc}.
The ${\mathcal O}(\alpha_S^2)$ term ${\mathcal A}_2$ consists of two parts, the
interference of two-loop diagrams with the Born amplitude and the interference  of
one-loop diagrams among themselves:
\begin{displaymath}
{\mathcal A}_2 = {\mathcal A}_2^{(2\times 0)} + {\mathcal A}_2^{(1\times 1)}\;.
\end{displaymath}
The  term ${\mathcal A}_2^{(1\times 1)}$ was derived in 
\cite{Korner:2005rg,Korner:2008bn,Anastasiou:2008vd,Kniehl:2008fd}. ${\mathcal
A}_2^{(2\times 0)}$, originating from the two-loop  diagrams, can be decomposed
according to color and flavor structures as follows:
\bea  {\mathcal A}_2^{(2\times 0)}  &=&  (N_c^2-1) 
\Biggl\{ N_c^3 A + N_c B 
 + \frac{1}{N_c} C
 + \frac{1}{N_c^3} D + N_c^2 N_l E_l + N_c^2 N_h E_h + N_l F_l + N_h F_h \nn\\
& &  \hspace*{15mm}
+ \frac{N_l}{N_c^2} G_l + \frac{N_h}{N_c^2} G_h 
+ N_c N_l^2 H_l + N_c N_h^2 H_h  
+ N_c N_l N_h H_{lh} + \frac{N_l^2}{N_c} I_l \nn\\
& & \hspace*{15mm}
 + \frac{N_h^2}{N_c} 
I_h  +  \frac{N_l N_h}{N_c} I_{lh}  \Biggr\} \, ,
\label{colstruc}
\eea
where $N_l$ and $N_h$ are the number of light- and heavy-quark flavors, respectively.
The coefficients $A,B,\ldots,I_{lh}$ in Eq.~(\ref{colstruc}) are functions of $s$,
$t$, and $m$, as well as of the dimensional regulator $\varepsilon$. These quantities
were calculated in  \cite{Czakon:2007wk} in the approximation $s,|t|,|u| \gg m^2$.
For a fully differential description of top quark pair production at NNLO, the
complete mass dependence of ${\mathcal A}_2^{(2\times 0)}$ is required; to date no
such a result is available.  Recently, a formula which allows to calculate the IR
poles of a generic two-loop QCD amplitude was derived
\cite{Becher:2009kw,Ferroglia:2009ep} and employed to obtain analytic expressions for
all of the IR poles  for the top-quark pair production \cite{Ferroglia:2009ii}. 

In this work, we provide an exact analytic expression for the leading colour
coefficient $A$ in Eq.~(\ref{colstruc}), which receives contributions from planar
Feynman diagrams only.

\section{Calculation\label{sec:calc}}


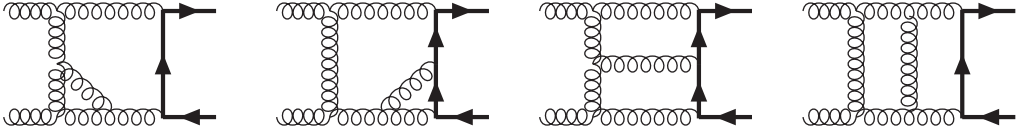
\begin{figure}
\vspace*{.5cm}
\[ \hspace*{-3mm}
\vcenter{
\hbox{\begin{picture}(0,0)(0,0)
\SetScale{1}
 \SetWidth{.5}
 \Gluon(-40,20)(-20,20){3}{4.5}
 \Gluon(-20,0)(-20,-20){3}{4.5}
 \Gluon(-20,20)(-20,0){3}{4.5}
 \Gluon(-20,-20)(-40,-20){3}{4.5}
 \Gluon(-20,20)(20,20){3}{7.5}
 \Gluon(-20,-20)(20,-20){3}{7.5}
 \Gluon(-20,0)(0,-18){3}{4.5}
 \SetWidth{1.6}
 \ArrowLine(20,20)(40,20)
 \ArrowLine(20,-20.5)(20,20.5)
 \ArrowLine(40,-20)(20,-20)
\end{picture} }
}
\hspace{3.5cm}
 \vcenter{
\hbox{\begin{picture}(0,0)(0,0)\SetScale{1}
 \SetWidth{.5}
 \Gluon(-40,20)(-20,20){3}{4.5}
 \Gluon(-20,20)(-20,-20){3}{9}
 \Gluon(-20,-20)(-40,-20){3}{4.5}
 \Gluon(-20,20)(20,20){3}{7.5}
 \Gluon(-20,-20)(20,-20){3}{7.5}
 \Gluon(0,-18)(20,0){3}{4.5}
 \SetWidth{1.6}
 \ArrowLine(20,20)(40,20)
 \ArrowLine(20,-20.5)(20,0)
 \ArrowLine(20,0)(20,20.5)
 \ArrowLine(40,-20)(20,-20)
\end{picture}}
}
\hspace{3.5cm}
 \vcenter{
\hbox{\begin{picture}(0,0)(0,0)
\SetScale{1}
 \SetWidth{.5}
 \Gluon(-40,20)(-20,20){3}{4.5}
 \Gluon(-20,20)(-20,0){3}{4.5}
 \Gluon(-20,0)(-20,-20){3}{4.5}
 \Gluon(-20,-20)(-40,-20){3}{4.5}
 \Gluon(-20,20)(20,20){3}{7.5}
 \Gluon(-20,-20)(20,-20){3}{7.5}
 \Gluon(-20,0)(20,0){3}{7.5}
 \SetWidth{1.6}
 \ArrowLine(20,20)(40,20)
 \ArrowLine(20,-20.5)(20,0)
 \ArrowLine(20,0)(20,20.5)
 \ArrowLine(40,-20)(20,-20)
\end{picture}}
}
\hspace{3.5cm}
 \vcenter{
\hbox{\begin{picture}(0,0)(0,0)
\SetScale{1}
 \SetWidth{.5}
 \Gluon(-40,20)(-20,20){3}{4.5}
 \Gluon(-20,20)(-20,-20){3}{9}
 \Gluon(-20,-20)(-40,-20){3}{4.5}
 \Gluon(-20,20)(20,20){3}{7.5}
 \Gluon(-20,-20)(20,-20){3}{7.5}
 \Gluon(0,18)(0,-18){3}{8.5}
 \SetWidth{1.6}
 \ArrowLine(20,20)(40,20)
 \ArrowLine(20,-20.5)(20,20.5)
 \ArrowLine(40,-20)(20,-20)
\end{picture}}
}
\]
\vspace*{.3cm}
\caption{\it Some of the two-loop planar box diagrams involved in the
calculation.}
\label{boxes}
\end{figure}


The package QGRAF~\cite{qgraf} generates 789 two-loop Feynman diagrams contributing
to the process $g  g\to t\bar t$ (considering one massless and one massive flavor).
Evaluating their color structures, we find that 300 of them contribute to  the
leading color coefficient in Eq.~(\ref{colstruc}). Since we use the covariant sum
over the polarizations of the incoming gluons, we have to consider additional 116
(out of 218) diagrams for the process initiated by ghosts.
We interfere the two-loop diagrams with the tree-level amplitude, and calculate the
color and Dirac traces before carrying out the integration over the loop momenta. The
resulting scalar loop integrals are reduced to a set of Master Integrals (MIs)
employing the technique based upon the Laporta algorithm \cite{Laportaalgorithm}. 
Then, the MIs can be evaluated analytically by means of the differential equation 
method \cite{DiffEq}.

Starting with the QGRAF output, the calculation of the interferences in terms of the
MIs was carried out with a parallelized {\tt C++} package: {\tt Reduze
2}~\cite{Reduze2}. In particular, this code provides a fully distributed variant of
the Laporta algorithm for the reduction\footnote{
Other reduction codes released publicly are
the {\tt Maple} package {\tt A.I.R.}~\cite{Anastasiou:2004vj},
the {\tt Mathematica} package {\tt FIRE}~\cite{Smirnov:2008iw}
and the {\tt C++} package {\tt Reduze}~\cite{Studerus:2009ye}.
}
of the loop integrals. The package employs {\tt GiNaC}~\cite{ginac} for the algebraic
manipulations.

The MIs needed for the calculation presented in this paper were already known  in the
literature 
\cite{Argeri:2002wz,Bonciani:2003te,Aglietti:2003yc,DK,Aglietti:2004tq,CGR,heavytolight,quarkloops,planarqq}.
In particular, all of the  two-loop four-point MIs encountered in the calculation of
the leading color coefficient in the gluon-fusion channel coincide with the ones
needed for the corresponding calculation in the quark-antiquark  annihilation channel
\cite{planarqq}, or can be obtained by the latter by replacing the Mandelstam
variable $t$ with $u$.
All of the MIs were calculated in the non-physical region $s<0$, where they are
real. The result in the physical region is then recovered by analytic continuation.

The transcendental functions appearing in the results are one- and two-dimensional
harmonic polylogarithms (HPLs)~\cite{HPLs,Vollinga,2dHPLs,Aglietti:2004tq} of maximum
weight four. All of the HPLs appearing in the analytic expression of the coefficient
$A$ can be evaluated numerically with arbitrary precision by employing the methods
and codes described  in~\cite{Vollinga}. A detailed list of the functional basis
employed in this calculation can be found in Appendix~A of~\cite{planarqq}.
Appendix~B in the same paper describes the expansion of this class of HPLs in the
threshold limit $\beta \to 0$, which requires some care. As expected, the sum of the bare
two-loop corrections (as well as the UV renormalized corrections) is symmetric with
respect to the exchange $t \leftrightarrow u$.

\section{Renormalization\label{sec:renorm}}

The renormalized QCD amplitude is obtained from the bare one by expanding
the following expression :
\bea
{\mathcal A}_{{\small \mbox{ren}}} & = & 
\prod_{n} Z^{1/2}_{{\tiny \mbox{WF}},n} \,
{\mathcal A}_{{\small \mbox{bare}}}
\left( \alpha_{S, {\small \mbox{bare}}} \to Z_{\alpha_S} \alpha_S \, ,
m_{{\small \mbox{bare}}} \to Z_m m \right) \, , \\
& = & Z_{{\tiny \mbox{WF}},g} \, Z_{{\tiny \mbox{WF}},t} \,
{\mathcal A}_{{\small \mbox{bare}}}
\left( \alpha_{S, {\small \mbox{bare}}} \to Z_{\alpha_S} \alpha_S \, ,
m_{{\small \mbox{bare}}} \to Z_m m \right) \, ,
\label{renM}
\eea
where $Z_{{\tiny \mbox{WF}},n}$ ($n=g,t$) are the external particle wave-function
renormalization factors, $\alpha_S$ is the renormalized coupling constant,
$Z_{\alpha_S}$ is the coupling constant renormalization factor, $m$ is the 
renormalized heavy-quark mass, and $Z_m$ is the mass renormalization factor.  In the
rest of the section we suppress the subscript ``$S$'' in $\alpha_S$.

We introduce the following auxiliary quantities:
\be
a_0 = \frac{\alpha_{\small \mbox{bare}}}{\pi}  \, ,
\qquad \mbox{and} \qquad
a = \frac{\alpha}{\pi}  \, .
\ee
By expanding the amplitude and the wave function renormalization factor in $a_0$ we
find:
\bea
{\mathcal A}_{{\small \mbox{ren}}}(\alpha_{\small \mbox{bare}}) &=&
    a_0 {\mathcal A}_0
  + a_0^2 {\mathcal A}_1
  + a_0^3 {\mathcal A}_2
  + {\mathcal O}(a_0^4) \, , \nn \\
Z_{{\tiny \mbox{WF}},n} & = & 1
  + a_0 \delta Z^{(1)}_{{\tiny \mbox{WF}},n}
  + a_0^2 \delta Z^{(2)}_{{\tiny \mbox{WF}},n}
  + {\mathcal O}(a_0^3) \, , \quad (n=g,t) \, , \nn \\
Z_m & = & 1
  + a_0 \delta Z^{(1)}_{m}
  + a_0^2 \delta Z^{(2)}_{m}
  + {\mathcal O}(a_0^3) \, .
  \label{exp1}
\eea
The relation between $a_0$ and $a$ is given by:
\be
a_0  =  a
  + a^2 \delta Z^{(1)}_{\alpha}
  + a^3\delta Z^{(2)}_{\alpha}
  + {\mathcal O}(a^4) \, . \label{exp2}
\ee
By employing  Eqs.~(\ref{exp1},\ref{exp2}) in Eq.~(\ref{renM}) we find
\bea
{\mathcal A}_{{\small \mbox{ren}}} & = & a {\mathcal A}_0
+ a^2 {\mathcal A}^{(1)} _{{\small \mbox{ren}}} +
a^3 {\mathcal A}^{(2)} _{{\small \mbox{ren}}} + \mathcal{O}(a^4) \, , \nn \\
{\mathcal A}^{(1)} _{{\small \mbox{ren}}} & = & {\mathcal A}_1 +
\left(  \delta Z^{(1)}_{{\tiny \mbox{WF}},t} 
      + \delta Z^{(1)}_{{\tiny \mbox{WF}},g} 
      + \delta Z^{(1)}_{\alpha} \right)  {\mathcal A}_0 
      - \delta Z_m^{(1)} 
          {\mathcal A}^{(\mbox{{\tiny {\tt mass CT}}}\!)}_0 
      \, , \nn \\
{\mathcal A}^{(2)} _{{\small \mbox{ren}}} & = &  {\mathcal A}_2 +
 \left(  \delta Z^{(1)}_{{\tiny \mbox{WF}},t} 
       + \delta Z^{(1)}_{{\tiny \mbox{WF}},g} 
       + 2 \delta Z^{(1)}_{\alpha}  \right) {\mathcal A}_1
+ \left(  \delta Z^{(1)}_{{\tiny \mbox{WF}},t} \delta Z^{(1)}_{{\tiny \mbox{WF}},g} 
         + 2 \delta Z^{(1)}_{{\tiny \mbox{WF}},t} \delta Z^{(1)}_{\alpha} 
    \right. \nn\\
& & \left.
         + 2 \delta Z^{(1)}_{{\tiny \mbox{WF}},g} \delta Z^{(1)}_{\alpha} 
         + \delta Z^{(2)}_{{\tiny \mbox{WF}},t}
         + \delta Z^{(2)}_{{\tiny \mbox{WF}},g}
         + \delta Z^{(2)}_{\alpha} 
\right) {\mathcal A}_0 - \delta Z_m^{(1)} 
          {\mathcal A}^{(\mbox{{\tiny {\tt mass CT}}}\!)}_1 \nn\\
& &
	  - \left(   \delta Z_m^{(2)} + 2 \, \delta Z^{(1)}_{\alpha} \delta Z_m^{(1)} 
	           + \delta Z^{(1)}_{{\tiny \mbox{WF}},t}  \delta Z_m^{(1)} 
            \right) {\mathcal A}^{(\mbox{{\tiny {\tt mass CT}}}\!)}_0 
	  + \left( \delta Z_m^{(1)} \right)^2 \, 
	  {\mathcal A}^{(\mbox{{\tiny {\tt 2 mass CT}}}\!)}_0 \, .
\label{exprenM} 
\eea
In the equations above, ${\mathcal A}_i$, $i=0,1,2$, represents the bare $i$-loop amplitude 
stripped of the factor $a$, ${\mathcal A}^{(\mbox{{\tiny {\tt mass CT}}}\!)}_0$ 
represents the amplitude obtained by adding a mass insertion to the virtual
top propagator in diagrams (b) and (c) of Fig.~\ref{figTree}. 
${\mathcal A}^{(\mbox{{\tiny {\tt mass CT}}}\!)}_1$ can be obtained by considering
all possible mass insertions in the one-loop diagrams. Finally, 
${\mathcal A}^{(\mbox{{\tiny {\tt 2 mass CT}}}\!)}_0$ is obtained by adding
two mass insertions to the virtual propagator in diagrams (b) and (c) of 
Fig.~\ref{figTree}. 

In this work we employ a mixed renormalization scheme in which the wave  functions
and the heavy-quark mass are renormalized on shell, while the  strong coupling
constant is renormalized in the $\overline{\mbox{MS}}$ scheme.
The explicit expressions of the one-loop renormalization factors in 
Eq.~(\ref{exprenM}) are:
\bea
\delta Z^{(1)}_{{\tiny \mbox{WF}},t} &=& C(\vep) \,
             \left(\frac{\mu^2}{m^2} \right)^{\vep} C_F
         \left( -\frac{3}{4 \ep} - \frac{1}{1-2 \ep}  \right) \, , \\
\delta Z^{(1)}_{{\tiny \mbox{WF}},g} &=& 0 \, , \\
\delta Z^{(1)}_{\alpha} &=& C(\vep) \,
\frac{e^{-\gamma \vep}}{\Gamma(1+\vep)} \left( -\frac{\beta_0}{2 \vep}\right) \, , \\
\delta Z^{(1)}_{m} &=& \delta Z^{(1)}_{{\tiny \mbox{WF}},t} \, ,
\label{onelct}
\eea
where $C(\vep) = (4 \pi)^{\vep} \Gamma(1+\vep)$,  $\beta_0 = (11/6) C_A  -(1/3) (N_l
+ N_h)$ and where $\gamma$ is the Euler-Mascheroni constant $\gamma \approx 
0.577216$.

Concerning the two-loop renormalization factors, we provide only the terms which 
contribute to the renormalization of the leading color coefficient. They are 
(see for instance~\cite{Melnikov:2000zc}):
\bea
\delta Z^{(2)}_{{\tiny \mbox{WF}},t} &=& C(\vep)^2 \left(\frac{\mu^2}{m^2} \right)^{2 \vep} \, C_F \Biggr[
C_F \left(\frac{9}{32\vep^2} + \frac{51}{64 \vep} + \frac{433}{128} -\frac{3}{2} \zeta(3)  
+ 6 \zeta(2) \ln{2} - \frac{39}{8} \zeta(2) \right) \nn \\
&& + C_A\left(-\frac{11}{32 \vep^2}-\frac{101}{64 \vep}-
\frac{803}{128} + \frac{3}{4} \zeta(3) - 3 \zeta(2) \ln{2} + \frac{15}{8} \zeta(2) \right)
 +
   {\mathcal O} \left( \ep\right)\Biggl] \, , \\
\delta Z^{(2)}_{{\tiny \mbox{WF}},g} &=& 0 \, , \\
\delta Z^{(2)}_{\alpha} &=& C(\vep)^2 \, \left( \frac{e^{-\gamma \vep}}{\Gamma(1+\vep)}
\right)^2 \frac{1}{4 \vep} \left[
\left(\frac{11}{6}\right)^2\frac{C_A^2}{\vep} - \frac{17}{12} C_A^2
 \right]  \, , \\
\delta Z^{(2)}_{m} &=&  C(\vep)^2 \left(\frac{\mu^2}{m^2} \right)^{2 \vep} \, C_F \Biggr[
C_F \left( \frac{9}{32 \vep^2} + \frac{45}{64 \vep} + \frac{199}{128} - \frac{15}{8} \zeta(2) 
          - \frac{3}{4} \zeta(3) +  3 \zeta(2) \ln{2} \right) \nn \\
&& + C_A \left( - \frac{11}{32 \vep^2} - \frac{91}{64 \vep} - \frac{605}{128} + \frac{1}{2} \zeta(2) 
          + \frac{3}{8} \zeta(3) - \frac{3}{2} \zeta(2) \ln{2} 
\right)
 +
   {\mathcal O} \left( \ep\right) \Biggl] 
\label{2lmrencoeff} \, .
\eea

\section{Results\label{sec:results}}

\begin{table} 
\begin{center}
\begin{tabular}{|c|c|c||c|c|c|c|c|}
\hline
& & & & & & & \\[-2.5ex]
$\dfrac{s}{m^2}$ & $\dfrac{t}{m^2}$ & $\dfrac{\mu}{m}$ &
  $1/\vep^4$ & $1/\vep^3$ & $1/\vep^2$ & $1/\vep$ &
  $\vep^0 $ \\[1.5ex]
\hline
\hline
$5$ & $-1.25$ & 1 &
  $10.749426$ & $18.693893$ & $-156.82372$ & $262.14826$ & $12.721807$ \\
\hline
$43$ & $-21$ & $1.7$ &
  $9.3642942$ & $-27.358589$ & $-41.372387$ & $305.80422$ & $-707.01281$ \\
\hline
$8.1$ & $-0.6$ & $2.1$ &
   $27.306074$ & $138.38466$ & $-381.11526$ & $-571.56293$ & $1859.6594$ \\
\hline
\end{tabular}
\end{center}
\caption{{\em Numerical values of the coefficients appearing in the $\varepsilon$
expansion of $A$ in the normalization that factors out the term 
$\bigl[ \left(4 \pi \right)^{\vep} e^{- \gamma \varepsilon} \bigr]^2 $.
} \label{tab:ben}}
\end{table}

\begin{figure}
\centering
\includegraphics[width=0.65\textwidth]{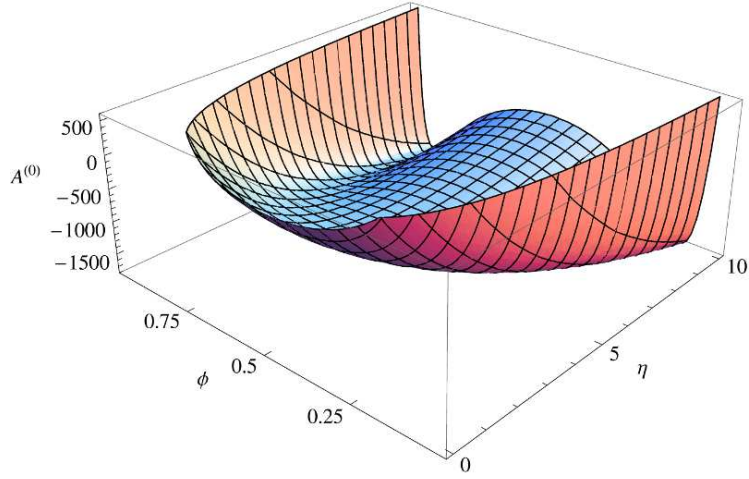} 
\caption{\it Finite part of the coefficient $A$ as a function of the variables
$\eta \equiv s/(4 m^2) - 1$ and $\phi \equiv -(t-m^2)/s$.
We employed the normalization that factors out the term 
$\bigl[ \left(4 \pi \right)^{\vep} e^{- \gamma \varepsilon} \bigr]^2 $
and set $\mu = m$.
}
\label{surface}
\end{figure}

\begin{figure}
\centering
\includegraphics[width=0.48\textwidth]{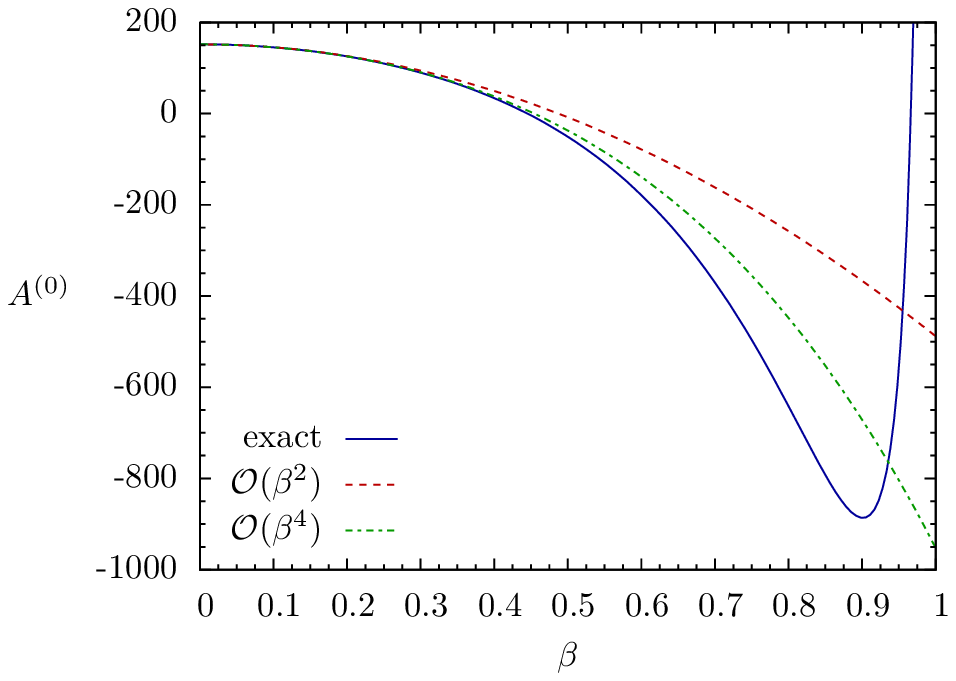}
\hfill
\includegraphics[width=0.48\textwidth]{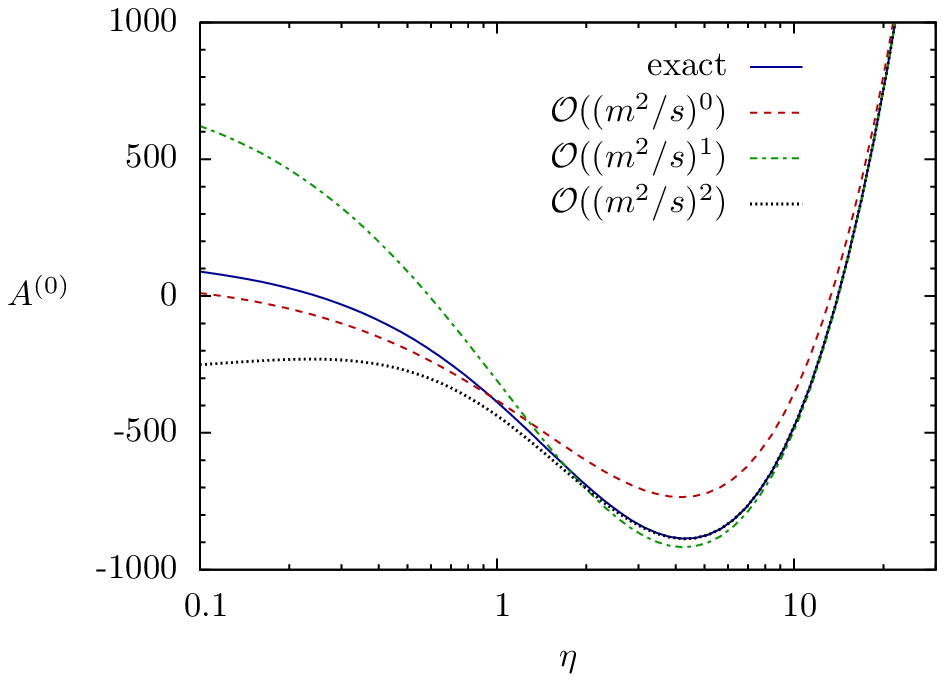} 
\caption{\it Left: finite part of the coefficient $A$ as a function of
$\beta \equiv \sqrt{1 - 4 m^2/s}$ for $\xi = 1/2$.
Right: finite part of the coefficient $A$ as a function of the 
variable $\eta \equiv s/(4 m^2) - 1$ for $\phi = 1/2$.
In both cases we used the normalization that factors out the term
$\bigl[ \left(4 \pi \right)^{\vep} e^{- \gamma \varepsilon} \bigr]^2 $
and set $\mu=m$.}
\label{expansions}
\end{figure}

The main result of the present paper is an analytic expression for the coefficient
$A$ in Eq.~(\ref{colstruc}) which retains the complete dependence on the variables
$t$, $u$, on the renormalization scale $\mu$, and on the top-quark mass $m$. The
analytic result is  too long to be explicitly presented in written form. To make it
available to the reader we include in the arXiv submission of this work a text file
with the complete result. The coefficients in Eq.~(\ref{colstruc}) still contain
infrared poles. This makes our result dependent on the choice of a global,
$\vep$-dependent normalization factor. We choose to present the coefficient $A$
factoring out an overall term
\be
C^2(\vep) = \bigl[\left(4 \pi \right)^{\vep} \Gamma(1+\vep) \bigr]^2 \, .
\label{Cep}
\ee
Another possible normalization, widely used in the literature, is the one in which 
a term 
\be
\tilde{C}^2(\vep) = \bigl[ \left(4 \pi \right)^{\vep} e^{- \gamma \varepsilon}
\bigr]^2 
\label{Cept}
\ee
is retained as an
overall factor (see for instance \cite{Ferroglia:2009ii} and \cite{Czakon:2008zk}).
Our result, therefore, can be compared to the results in \cite{Ferroglia:2009ii,Czakon:2008zk} 
by multiplying it by a coefficient
\be
e^{2 \gamma \varepsilon} \, \Gamma^2(1+\varepsilon) = 
1 + \zeta(2) \varepsilon^2 - \frac{2}{3}\zeta(3) \varepsilon^3 
+ \frac{7}{10} \zeta(2)^2\varepsilon^4 
+ {\mathcal O}\left(\varepsilon^5\right)\, .
\ee
While the formulae in the text and the supplied files are given in the normalization
defined by Eq.~(\ref{Cep}), for ease of comparison with other results in the 
literature we present in Table~\ref{tab:ben} and in 
Figures~\ref{surface}--\ref{expansions} a few benchmark points and plots in the
normalization of  Eq.~(\ref{Cept}).

We provide a code which numerically evaluates the analytic expression of the leading
color coefficient for arbitrary values of the mass scales involved in the
calculation. The code is written in {\tt C++} and uses the package for the evaluation
of multiple polylogarithms within {\tt GiNaC}~\cite{Vollinga}.

Our result was cross checked by comparing it with the partial results already
available in the literature. By expanding the analytic expression of the coefficient
$A$ in the limit $s,|t|,|u| \gg m^2$ and neglecting terms suppressed by positive
powers of $m^2/s$ for $t/s = \mathit{const}$, we obtained the result published
in \cite{Czakon:2007wk}.
For the IR poles of the coefficient $A$ we find numerical agreement with
the results of \cite{Ferroglia:2009ii}.

In Fig.~\ref{surface} the finite part of $A$ is plotted as a surface depending 
on the variables $\eta$ and $\phi$, defined as
\be
\eta = \frac{s}{4 m^2} - 1 \, , \quad \phi = - \frac{t -m^2}{s} \,,
 \quad \frac{1}{2}\left(1-\sqrt{\frac{\eta}{1+\eta}} \right) \le \phi
 \le \frac{1}{2}\left(1+\sqrt{\frac{\eta}{1+\eta}} \right).
\ee

Finally, it is possible to expand the expression for $A$ for values of the center of mass
energy close to the production threshold. We define
\be
 \beta = \sqrt{1-\frac{4 m^2}{s}} \, , \qquad
 \xi = \frac{1-\cos\theta}{2} \, , \qquad
 L_\mu = \ln\left(\frac{\mu^2}{m^2} \right) \, , 
 \label{deff}
\ee
where $\theta$ is the scattering angle in the partonic center of mass frame.
Keeping $\xi = \mathit{const}$, we expand our results in powers of the heavy
quark velocity $\beta$ up to terms of order $\beta^5$ included.
The coefficients of this expansion contain transcendental
constants which originate from one- and two-dimensional HPLs  evaluated at $x=1$.
Since we did not find a satisfactory analytical representation for all of these
constants, in the formulae below we provide  the coefficients of the expansion in a
numerical form. We find:
\be
A(\beta,\xi) = \frac{A^{( -4)}(\beta,\xi)}{\vep^4} \! 
                 + \! \frac{A^{( -3)}(\beta,\xi)}{\vep^3} \!
                 + \! \frac{A^{( -2)}(\beta,\xi)}{\vep^2} \!
     + \! \frac{A^{( -1)}(\beta,\xi)}{\vep} 
     + A^{(0)}(\beta,\xi) 
+ {\mathcal O}\left(\vep\right)
\ee
with
\begin{eqnarray}
A^{( -4)} &=&
 8
 + \left[24 - 32 \xi(1-\xi)
   \right] \beta^2
 + \left[32 - 128 \xi(1-\xi) - 256 (\xi(1-\xi))^2
   \right] \beta^4 \nn\\
& & 
 + {\mathcal O}(\beta^6) \, , \\
A^{( -3)} &=&
 13.1526 + 16 L_\mu
 + \left[71.4579+ 48 L_\mu - (84.6105 + 64 L_\mu) \xi(1-\xi)
   \right] \beta^2
 \nn\\ & &
 + \left[103.277+ 64 L_\mu - (487.775 + 256 L_\mu) \xi(1-\xi)
 \right. \nn\\ & &\quad \left.
                           - (1850.22 + 512 L_\mu) (\xi(1-\xi))^2)
   \right] \beta^4 
 + {\mathcal O}(\beta^6) \, , \\
A^{( -2)} &=&
 -155.954 - 32.3614 L_\mu + 16 L_\mu^2
 \nn\\ & &
 - \left[ 388.711 + 33.0843 L_\mu - 48 L_\mu^2
         - (1148.69+ 65.4457 L_\mu - 64 L_\mu^2) \xi(1-\xi)
   \right] \beta^2
 \nn\\ & &
 - \left[   461.697 + 28.1123 L_\mu - 64 L_\mu^2
         - (2976.85 - 36.884 L_\mu - 256 L_\mu^2) \xi(1-\xi)
 \right. \nn\\ & &\quad \left.
         + ( 2814.61 + 1823.1 L_\mu + 512 L_\mu^2) (\xi(1-\xi))^2
   \right] \beta^4
 + {\mathcal O}(\beta^6) \, , \\
A^{( -1)} &=&
  224.144- 191.469 L_\mu - 61.6948 L_\mu^2 + 10.6667 L_\mu^3
 \nn\\ & &
  + \left[  117.543- 592.101 L_\mu - 121.084 L_\mu^2 + 32 L_\mu^3
 \right. \nn\\ & &\quad \left.
        - ( 774.295 - 1463.61 L_\mu - 182.779 L_\mu^2 + 42.6667 L_\mu^3) \xi(1-\xi)
    \right] \beta^2 \nn\\
& &
  - \left[ 92.4819 + 744.745 L_\mu + 145.446 L_\mu^2 - 42.6667 L_\mu^3
 \right. \nn\\ & &\quad \left.
        + ( 553.64 - 4574.2 L_\mu - 432.449 L_\mu^2 + 170.667 L_\mu^3) \xi(1-\xi)
 \right. \nn\\ & &\quad \left.
        - (17463.5 \! + \! 2328.24 L_\mu \! - \! 884.435 L_\mu^2 \! 
	- \! 341.333 L_\mu^3) (\xi(1-\xi))^2
    \right] \beta^4 \nn\\
& &  
  + \! {\mathcal O}(\beta^6) , \\
A^{(0)} &=&
  403.869+ 787.434 L_\mu - 77.4706 L_\mu^2 - 50.9076 L_\mu^3 + 5.33333 L_\mu^4
 \nn\\ & &
  + \left[1627.76+ 1320.97 L_\mu -338.107 L_\mu^2 - 110.056 L_\mu^3 + 16 L_\mu^4
 \right. \nn\\ & &\quad \left.
       - ( 8363.19 + 4515.06 L_\mu - 831.618 L_\mu^2 - 160.964 L_\mu^3 
       + 21.3333 L_\mu^4) \xi(1-\xi)
    \right] \beta^2
 \nn\\ & &
 + \left[ 1826.76+ 1164.29 L_\mu - 440.308 L_\mu^2 - 136.075 L_\mu^3 
 + 21.3333 L_\mu^4
 \right. \nn\\ & &\quad \left.
      - ( 22884.9 + 10052.4 L_\mu - 3024.01 L_\mu^2 - 444.744 L_\mu^3 
      + 85.3333 L_\mu^4) \xi(1-\xi)
 \right. \nn\\ & &\quad \left.
      + (50837.4 \! + \! 35185.8 L_\mu \! + \! 4586.08 L_\mu^2 \! 
      - \! 276.734 L_\mu^3 \! - 
        \!  170.667 L_\mu^4) (\xi(1-\xi))^2
   \right] \beta^4
 \nn\\ & &
 + {\mathcal O}(\beta^6) \, .
\end{eqnarray}
We observe that the dependence on $\beta$ and on $\xi$ in the formulas above is 
polynomial. Moreover, $\beta$ enters only via powers of $\beta^2$ and $\xi$ only via
powers of $\xi(1-\xi)$. The latter dependence explicitely reflects the symmetry of $A$
under exchange of forward and backward directions, $\xi \to 1-\xi$. All of the
logarithmic terms  $\ln{\beta}, \, \ln{\xi}, \, \ln{(1-\xi)}, \, \ln{(1-2\,\xi)}, \,
\ldots$, which are indeed present in the expansion of individual HPLs, cancel out in
the final  expressions. 
Thus, the coefficient $A$ is finite at threshold.
The expansion presented here could be used in the future for the calculation of
logarithmically enhanced leading colour terms near the $t \bar{t}$ production
threshold. 

In Fig.~\ref{expansions}, we show the comparison between the exact expression
of the coefficient $A$ and the expansions in the two regimes:
threshold expansion on the left hand side and
small-mass expansion on the right hand side.
For the small-mass expansion we consider the limit $m^2/s \to 0$ for
$\phi=-(t-m^2)/s = \mathit{const}$ and neglect terms suppressed
by powers of $m^2/s$ larger than a given order.
This prescription preserves the symmetry of the result under exchange of forward
and backward directions, $\phi \to 1-\phi$, for a given order of the expansion.

With the arXiv submission of this paper, we provide algebraic expressions and
numerical implementations of the threshold expansion up to order $\beta^5$
included and of the small mass expansion up to order $(m^2/s)^2$ included.

\section{Conclusions and Outlook\label{sec:conc}}

In this paper, we presented the analytic calculation of the two-loop leading
color corrections to the heavy-quark production matrix element in the gluon fusion
channel.  The diagrams required to calculate this coefficient are
all planar. The result presented here retains the exact heavy-quark mass
dependence; no assumptions on the hierarchy between the mass scales involved in
the problem was made. 
The formula for the coefficient $A$ in Eq.~(\ref{colstruc}), which was obtained 
in this work, was validated against analytic results valid in
the small-mass region~\cite{Czakon:2007wk}. In addition, we found numerical agreement 
with the exact analytic expression for the IR poles which was obtained 
in \cite{Ferroglia:2009ii}.

Our result represents a gauge invariant sub-set of the full two-loop 
corrections to the partonic process $g g \to t \bar t$.
In order to complete the analytic calculation of the two-loop corrections,
it is necessary to calculate all of the fermionic diagrams, and the non-planar
gluonic diagrams.
A large part of the non-leading color
coefficients in Eq.~(\ref{colstruc}) can be calculated within the same
calculational framework employed here. However, it is known that some of the 
two-loop diagrams appearing in the gluon fusion channel cannot be expressed in
terms the HPLs functional basis. For example, some of the diagrams with a
closed heavy-quark loop involve a ``sunrise''-type subtopology with three
equal massive propagators and an external momentum which is not on the mass
shell of the internal propagators. Such a three-propagator graph can be written
only in terms of elliptic integrals \cite{Laporta:2004rb}. The use of numerical 
\cite{Czakon:2008zk} or seminumerical \cite{Pozzorini:2005ff,Aglietti:2007as} 
methods could thus be unavoidable in the evaluation of these diagrams.

In order to obtain NNLO predictions for the total $t \bar t$ production cross
section and for differential distributions, it is necessary to combine the two-loop
virtual corrections with the already  available one-loop corrections to the $t\bar
t$+(1~parton) process and with the tree-level matrix elements for the process $t\bar
t$+(2~partons)~\cite{Dittmaier:2007wz, Bevilacqua:2010ve, Melnikov:2010iu}.  These
diagrams with additional partons in the final state contribute to infrared-divergent
configurations where up to two partons can become unresolved. Their implementation
requires the application of a NNLO subtraction method. The methods presently
available~\cite{secdec,ant,cg} do not provide yet the full counterterms for a $2\to 2$
hadronic process involving  massive partons. Two subtraction methods for the NNLO
calculation of the top-quark pair production cross section were outlined recently
\cite{Czakon:2010td,Anastasiou:2010pw}.  

In the light of these advancements, the calculation of the full NNLO corrections
to top quark pair production in hadronic collisions is gradually
becoming feasible.

\subsection*{Acknowledgments}

Part of the algebraic manipulations of this work were done using FORM \cite{FORM}.\\
R.~B. wishes to thank the Center for Theoretical Physics of New York City 
College of Technology for the kind hospitality during a part of this work.
A.~v.~M. wishes to thank the Institute for Theoretical Physics of the University
of Bielefeld for the kind hospitality during a part of this work and
Y.~Schr\"oder for stimulating discussions.\\
The work of R.~B. is supported by the Theory-LHC-France initiative of 
CNRS/IN2P3 and partly funded by the Center for Theoretical Physics of New 
York City College of Technology. T.~G.\ and A.~v.~M.\ are supported by the Schweizer Nationalfonds
(grants 200020-126691 and 200020-124773).
The work of C.~St. was supported by the Deutsche Forschungsgemeinschaft 
(DFG SCHR 993/2-1).


\end{document}